\documentclass{llncs}
\usepackage{graphicx}
\usepackage{color}

\usepackage{mathrsfs}         

\newcommand{\SAT}[0]{\ensuremath{\textsc{Sat}}}
\newcommand{\PSAT}[0]{\ensuremath{\textsc{Planar 3\mbox{-}Sat}}}
\newcommand{\QP}[0]{\ensuremath{\textsc{Queens}}}

\newcommand{\NP}[0]{\ensuremath{\mathsf{NP}}}
\newcommand{\Pspace}[0]{\ensuremath{\mathsf{Pspace}}}
\newcommand{\Exptime}[0]{\ensuremath{\mathsf{Exptime}}}

\begin{document}

\title{On the Complexity of a Derivative Chess Problem
}
\author{Barnaby Martin\thanks{This paper is publicly available at http://arxiv.org/abs/cs.CC/0701049.}}
\institute{Department of Computer Science, University of Durham,\\
  Science Labs, South Road, Durham DH1 3LE, U.K.}
\maketitle

\begin{abstract}
We introduce \QP, a derivative chess problem based on the classical $n$-queens problem. We prove that \QP\ is \NP-complete, with respect to polynomial-time reductions.
\end{abstract}

\section{Introduction}

There is a considerable literature on the complexity of various games and puzzles. In the case of two player games, the complexity question is usually as to whether the first player has a winning strategy. Since the classical games are frequently finite, it is generally the case that this complexity is actually zero. For example, in classical $8 \times 8$ Chess, it is known that one of the following holds (although it is not yet known which): either white or black has a winning strategy or black may force stalemate. Therefore, to avoid considering only single instances, it is usual to consider generalised versions of these games. For example, Generalised Chess takes as input an $n \times n$ board, for some $n$, on which are placed any quantity of white and black pieces (subject to there being exactly one king of each colour). The problem is whether or not white has a winning strategy from this starting position. The complexity of this problem is known to be \Exptime-complete \cite{ChessExptime}. In fact many of these generalised two player games attain complexities of \Pspace\ or \Exptime. Further examples include that of Generalised Checkers, also \Exptime-complete \cite{CheckersExptime}, and Generalised Othello, \Pspace-complete \cite{OthelloPspace}. We will be more interested in one player games, which would more naturally be called puzzles. The complexity question here is usually as to whether or not there exists a solution to the puzzle. Such problems are prime candidates for \NP, and this will certainly be the case so long as a correct solution is easily verifiable. Many of these puzzles turn out to be \NP-complete: for example, Cross Sum (Kakuro) \cite{SetaSeniorThesis}, Crossword \cite{GareyJohnson}, Latin Square \cite{LatinSquareNP} and Sudoku \cite{YatoSeta}.      

The classical \emph{$n$-queens problem} considers an $n \times n$ chessboard and the problem as to whether $n$ queens may be placed thereon with no queen in a position to take any other. If this were to be considered as the problem, the input would simply be the number $n$, and the complexity would be zero space: it is known to have a solution for all $n\geq1$ other than $2$ and $3$ \cite{Madachy}. We instead consider a natural variant of this problem, \QP, in which the input is a number $m$ together \nopagebreak with a $k \times k'$ board whose squares are either black or white. The black squares are \nopagebreak to be interpreted as walls through which a queen may not pass: in this respect the board is somewhat different from a normal chess board, and more akin to a crossword grid. The problem is whether we may place $m$ queens on the board such that no queen is in a position to take any other. The contribution of this paper is to prove this problem \NP-complete, with respect to polynomial-time reductions. 

\section{Related Work}

The method of reduction through the construction of certain gadgets is ubiquitous in the theory of \NP-completeness. The method used in this paper is in the spirit of those  used to prove the \NP-completeness of Cross Sum \cite{SetaSeniorThesis} and Reflections \cite{ReflectionsNP}.

\section{Preliminaries}

The problem of \emph{propositional satisfiability}, \SAT, takes as input a collection of clauses, where each clause is a collection of literals, and each literal is a propositional variable or its negation. The input is a yes-instance iff all of the clauses, each seen as a disjunction of their literals, are simultaneously satisfiable by some truth assignment to the propositional variables. The \emph{incidence graph} of an instance of \SAT\ is formed by taking each of the variables and each of the clauses as vertices, with an edge connecting a variable to any clause in which it features. The problem \PSAT\ is the restriction of \SAT\ to inputs in which clauses have maximum size $3$ and the incidence graph is planar. It is known that \PSAT\ is \NP-complete \cite{Lichtenstein}.

We now define the problem \QP. The input is a $k \times k'$ board $\mathscr{G}$, each of whose squares is either black or white, and a tariff $m$. Recall that in chess a queen can move any distance in a horizontal, vertical or diagonal fashion. In contrast to a real chess board, we consider our black squares to be walls through which a queen can not move; she may therefore only sit on or threaten the white squares. An assignment of queens to the board is called \emph{legitimate} if all queens are placed on white squares and no queen is in a position to take any other. The input is a yes-instance if there is a legitimate assignment of $m$ queens to the board. Plainly the problem \QP\ is in \NP: we may guess an assignment of $m$ queens to $\mathscr{G}$ and readily verify that it is legitimate. The remainder of this paper concerns itself with proving that the problem is also \NP-hard. Note that the only flavour of reduction we consider is the many-to-one, polynomial-time reduction. For more on this, and the concept of \NP-completeness, see \cite{GareyJohnson}.

\section{Basic Constructions}

We will prove the \NP-hardness of \QP\ via a reduction from \PSAT. The properties of the following gadgets will be at the heart of this construction. The clause gadget is the following sub-board.

\hspace{4.5cm} \includegraphics{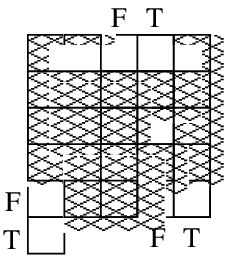}

\noindent The clause gadget is such that the maximum number of queens that may be legitimately placed within it is five, and that this is achievable iff one queen is placed in each of the three exterior double squares and not all of the three squares labelled $F$ contain queens. In this manner, the three exterior double squares naturally simulate the three literals of a clause in an instance of \PSAT. The eight possible truth assignments to a three-literalled clause are illustrated below. The first seven may be extended so that the sub-board holds $5$ queens; in these cases the drawn extension is not necessarily unique. The last -- three falses -- can only be extended to hold $4$ queens.

\hspace{1cm} \includegraphics{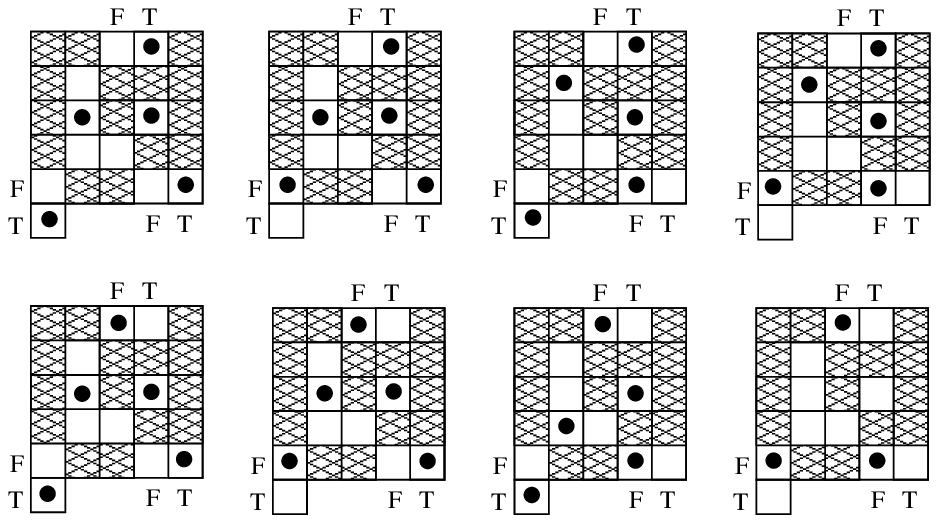}

\noindent All that remains is to connect like variables in different clauses, forcing consistency of the valuation of that variable whilst respecting that the variable may occur with different polarity within different clauses. To do this we use various combinations of the following gadgets (in each case there are exactly two legitimate maximal coverings, and these are illustrated with black and white circles, respectively).  

\hspace{2.3cm} \includegraphics{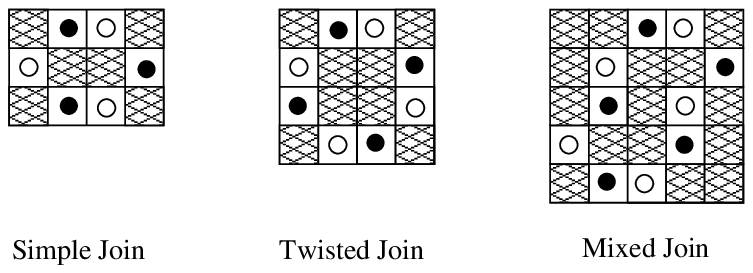}

\noindent It is not the case that these gadgets can be connected to one another willy-nilly whilst preserving their desired properties. For example, a Twisted Join and a Simple Join may not be connected to one another as shown in (ii) of the following diagram and be expected to legitimately hold $4+3-1=6$ (the minus $1$ is for the overlap) queens in precisely two ways. (Although the hybrid gadget (ii) \emph{can} hold $6$ queens in one way; and there are other ways in which a Twisted Join and a Simple Join may be connected in order that they can hold $6$ queens in precisely two ways.) However, certain combinations  do preserve the desired properties: for example the marriage of a Mixed Join and a Simple Join (respectively, two Twisted Joins), as in (i) (respectively, (iii)) results in a gadget that legitimately holds at most $5+3-1=7$ (respectively, $4+4=8$) queens, and this is achievable in precisely two ways.

\hspace{2cm} \includegraphics{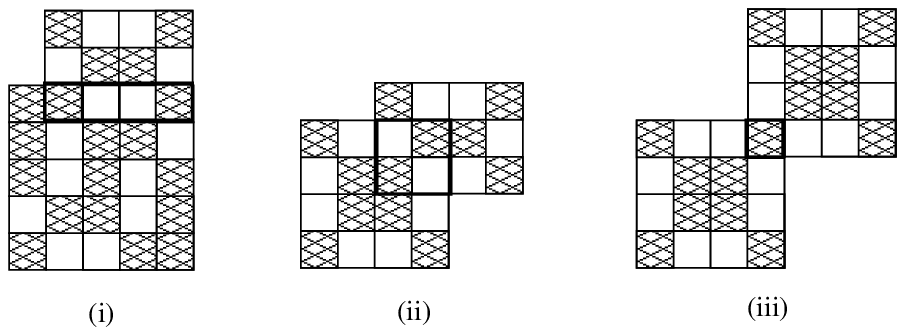}

\noindent The following lemmas use such suitable marriages of our gadgets. Their proofs are a matter of simple verification. In the case of the Clause Sub-board, it should be noted that a copy of the previously-introduced clause gadget lurks at its heart, and that the surrounding ephemera merely serve to propagate the valuation to the edges.

\begin{lemma}[Clause Sub-board]
The following $16 \times 16$ sub-board has the property that it can legitimately hold at most $21$ queens, and this is achievable iff one queen is placed in each of the three exterior double squares and not all of the three squares labelled $F$ contain queens.

\hspace{3.5cm}  \includegraphics{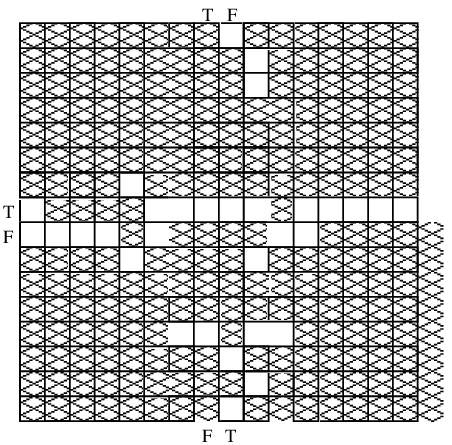}
\end{lemma}

\begin{lemma}[Variable Sub-board]
The following $16 \times 16$ sub-board has the property that it can legitimately hold at most $20$ queens, and this is achievable only in the two ways depicted.

\hspace{1cm} \includegraphics{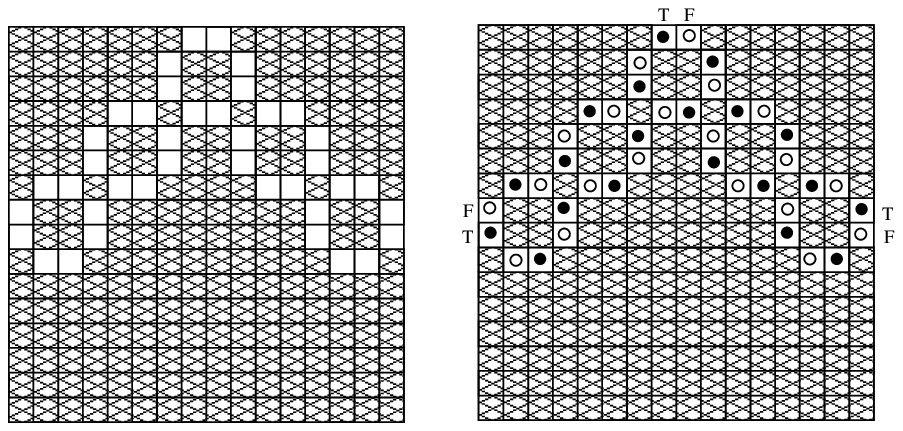}
\end{lemma}

\begin{lemma}[Turn Sub-board]
The following $16 \times 16$ sub-board has the property that it can legitimately hold at most $12$ queens, and this is achievable only in the two ways depicted.

\hspace{1cm} \includegraphics{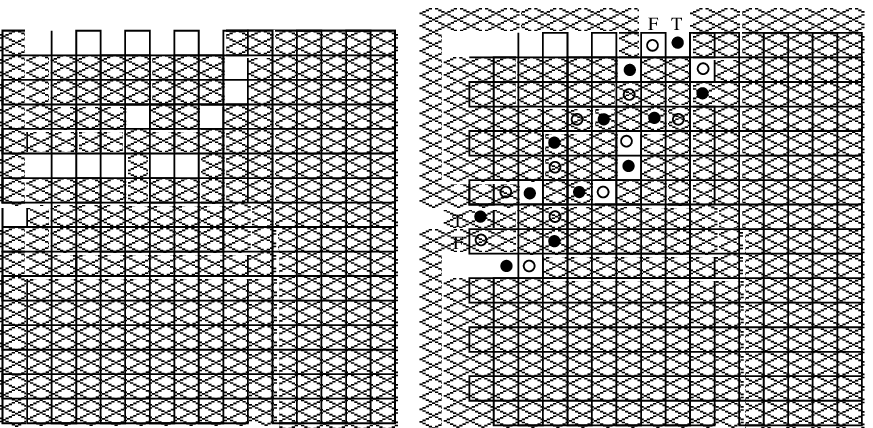}
\end{lemma}

\begin{lemma}[Join FM Sub-board]
The Join FM (Female-Male) Sub-board has the property that it can legitimately hold at most $16+1$ queens, including one on its protuberence, and this is achievable only in the two ways depicted. (Note that the $16 \times 16$ body can never hold more than $16$ queens, even if the protuberence remains empty.)

\hspace{1cm} \includegraphics{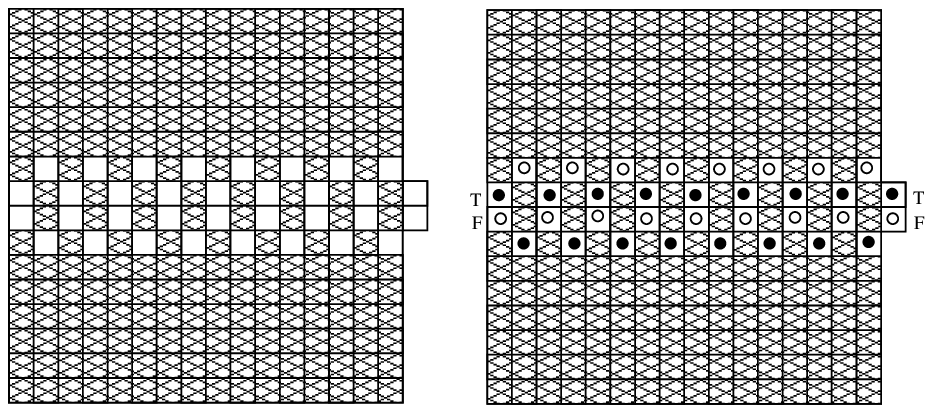}
\end{lemma}

\begin{lemma}[Join MM Sub-board]
The Join MM (Male-Male) Sub-board has the property that it can legitimately hold at most $17+2$ queens, including two on its protuberences, and this is achievable only in the two ways depicted. (Note that the $16 \times 16$ body can never hold more than $17$ queens, even if the protuberences remain empty.)

\hspace{1cm} \includegraphics{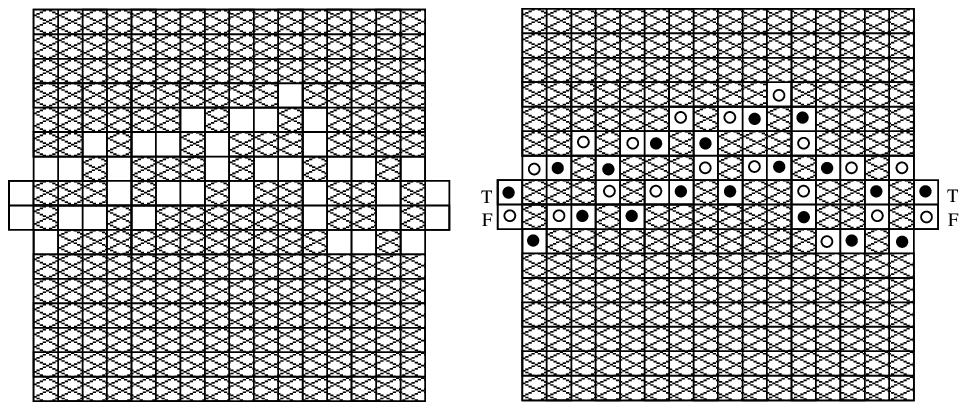}
\end{lemma}

\begin{lemma}[Join-Switch MM Sub-board]
The Join-Switch MM (Male-Male) Sub-board has the property that it can legitimately hold at most $16+2$ queens, including two on its protuberences, and this is achievable only in the two ways depicted. (Note that the $16 \times 16$ body can never hold more than $16$ queens, even if the protuberences remain empty.)

\hspace{1cm} \includegraphics{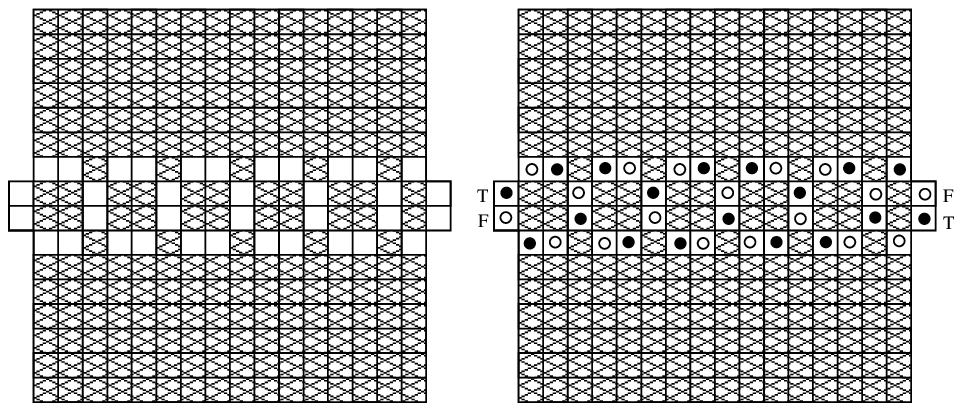}
\end{lemma}

We will use the following shorthands (and their rotations) for the sub-boards defined (the Black Sub-board is a $16 \times 16$ board with all squares black).

\vspace{0.5cm}
\hspace{-.2cm} \includegraphics{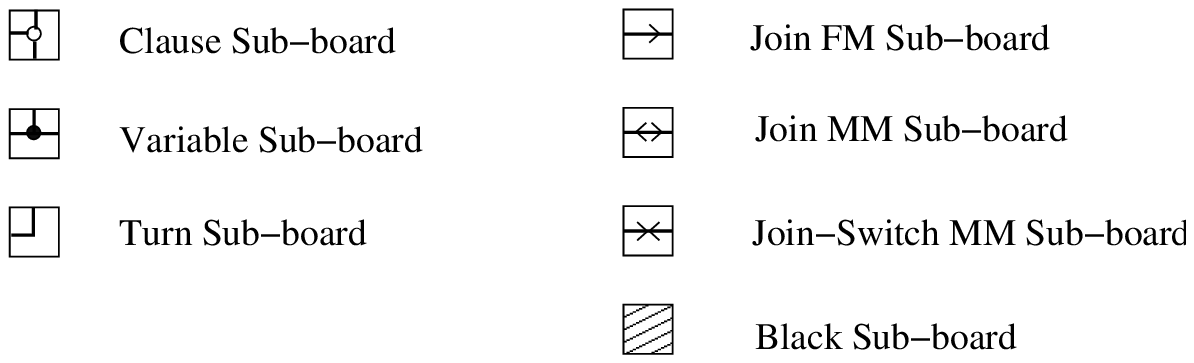}
\vspace{0.5cm}

\noindent It should be noted that in all of the sub-boards given, other than the Clause Sub-board, the choice of labelling $T$ and $F$ was arbitrary, and is somewhat artificial. In the case of the Variable Sub-board (as well as the Clause Sub-board) we will use only the labelling given. For the Turn and assorted Join sub-boards we may sometimes choose the converse labelling, although, of course, the sub-board itself remains the same.

\begin{theorem}[The main result]
The problem \QP\ is \NP-complete, with respect to polynomial-time reductions.
\end{theorem}
\begin{proof}
Membership of \NP\ is transparent -- we guess an assignment of the queens and verify its legitimacy. To prove \NP-hardness, we reduce from the problem \PSAT. As the reduction is explained it will be illustrated with a running example. Given an input $\Phi$ of \PSAT, we will construct a board $\mathscr{G}$ and tariff $m$ such that $m$ queens can be placed legitimately on $\mathscr{G}$ iff $\Phi$ is satisfiable. If the number of clauses plus the number of variables in $\Phi$ is $n$, then the size of $\mathscr{G}$ and $m$ will be $O(n^4)$, and, furthermore, the reduction will be computable in polynomial-time.  

From an input $\Phi$ of \PSAT\ we will first build some planar embedding $G_\Phi$ of its incidence graph. It is known this can be done in linear time \cite{embeddingplanar}. Our example instance will be
\[
\begin{array}{ccccc}
(\neg v_1), & (v_1,\neg v_2), & (v_1,v_2,v_3), & (v_1,\neg v_3,v_4), & (v_1,\neg v_2,\neg v_3).
\end{array}
\]
from which we may generate the planar embedding

\vspace{.5cm}
\hspace{.5cm} \begin{picture}(0,0)%
\includegraphics{planarembedding221.pstex}%
\end{picture}%
\setlength{\unitlength}{3947sp}%
\begingroup\makeatletter\ifx\SetFigFont\undefined%
\gdef\SetFigFont#1#2#3#4#5{%
  \reset@font\fontsize{#1}{#2pt}%
  \fontfamily{#3}\fontseries{#4}\fontshape{#5}%
  \selectfont}%
\fi\endgroup%
\begin{picture}(4555,1926)(274,-1126)
\put(3851,-161){\makebox(0,0)[lb]{\smash{{\SetFigFont{10}{12.0}{\rmdefault}{\mddefault}{\updefault}{\color[rgb]{0,0,0}$c_4$}%
}}}}
\put(4814,-149){\makebox(0,0)[lb]{\smash{{\SetFigFont{10}{12.0}{\rmdefault}{\mddefault}{\updefault}{\color[rgb]{0,0,0}$c_5$}%
}}}}
\put(289,-236){\makebox(0,0)[lb]{\smash{{\SetFigFont{10}{12.0}{\rmdefault}{\mddefault}{\updefault}{\color[rgb]{0,0,0}$c_1$}%
}}}}
\put(1201,-224){\makebox(0,0)[lb]{\smash{{\SetFigFont{10}{12.0}{\rmdefault}{\mddefault}{\updefault}{\color[rgb]{0,0,0}$c_2$}%
}}}}
\put(2101,-236){\makebox(0,0)[lb]{\smash{{\SetFigFont{10}{12.0}{\rmdefault}{\mddefault}{\updefault}{\color[rgb]{0,0,0}$c_3$}%
}}}}
\put(2839,-186){\makebox(0,0)[lb]{\smash{{\SetFigFont{10}{12.0}{\rmdefault}{\mddefault}{\updefault}{\color[rgb]{0,0,0}$v_4$}%
}}}}
\put(3326,-724){\makebox(0,0)[lb]{\smash{{\SetFigFont{10}{12.0}{\rmdefault}{\mddefault}{\updefault}{\color[rgb]{0,0,0}$v_3$}%
}}}}
\put(2689,-1061){\makebox(0,0)[lb]{\smash{{\SetFigFont{10}{12.0}{\rmdefault}{\mddefault}{\updefault}{\color[rgb]{0,0,0}$v_2$}%
}}}}
\put(2401,651){\makebox(0,0)[lb]{\smash{{\SetFigFont{10}{12.0}{\rmdefault}{\mddefault}{\updefault}{\color[rgb]{0,0,0}$v_1$}%
}}}}
\end{picture}%

\vspace{.5cm}

\noindent where the light circles represent the clauses and the dark circles the variables. Since we will want to represent this planar graph in a grid, we can not handle vertices of degree greater than three. It is possible that some variable vertices may have a greater degree $d$, $3 < d \leq n$, so we replace such vertices with $d -2$ colinear `subvariable' vertices, in the manner now illustrated, to obtain the planar graph $G'_\Phi:=$

\vspace{.5cm}
\hspace{.5cm} \begin{picture}(0,0)%
\includegraphics{planarembedding331.pstex}%
\end{picture}%
\setlength{\unitlength}{3947sp}%
\begingroup\makeatletter\ifx\SetFigFont\undefined%
\gdef\SetFigFont#1#2#3#4#5{%
  \reset@font\fontsize{#1}{#2pt}%
  \fontfamily{#3}\fontseries{#4}\fontshape{#5}%
  \selectfont}%
\fi\endgroup%
\begin{picture}(4546,2033)(274,-1253)
\put(289,-413){\makebox(0,0)[lb]{\smash{{\SetFigFont{10}{12.0}{\rmdefault}{\mddefault}{\updefault}{\color[rgb]{0,0,0}$c_1$}%
}}}}
\put(1200,-385){\makebox(0,0)[lb]{\smash{{\SetFigFont{10}{12.0}{\rmdefault}{\mddefault}{\updefault}{\color[rgb]{0,0,0}$c_2$}%
}}}}
\put(2138,-378){\makebox(0,0)[lb]{\smash{{\SetFigFont{10}{12.0}{\rmdefault}{\mddefault}{\updefault}{\color[rgb]{0,0,0}$c_3$}%
}}}}
\put(2851,-356){\makebox(0,0)[lb]{\smash{{\SetFigFont{10}{12.0}{\rmdefault}{\mddefault}{\updefault}{\color[rgb]{0,0,0}$v_4$}%
}}}}
\put(2625,-1182){\makebox(0,0)[lb]{\smash{{\SetFigFont{10}{12.0}{\rmdefault}{\mddefault}{\updefault}{\color[rgb]{0,0,0}$v_2$}%
}}}}
\put(3302,-871){\makebox(0,0)[lb]{\smash{{\SetFigFont{10}{12.0}{\rmdefault}{\mddefault}{\updefault}{\color[rgb]{0,0,0}$v_3$}%
}}}}
\put(1976,526){\makebox(0,0)[lb]{\smash{{\SetFigFont{10}{12.0}{\rmdefault}{\mddefault}{\updefault}{\color[rgb]{0,0,0}$v_1$}%
}}}}
\put(3062,526){\makebox(0,0)[lb]{\smash{{\SetFigFont{10}{12.0}{\rmdefault}{\mddefault}{\updefault}{\color[rgb]{0,0,0}$v''_1$}%
}}}}
\put(4805,-307){\makebox(0,0)[lb]{\smash{{\SetFigFont{10}{12.0}{\rmdefault}{\mddefault}{\updefault}{\color[rgb]{0,0,0}$c_5$}%
}}}}
\put(3853,-328){\makebox(0,0)[lb]{\smash{{\SetFigFont{10}{12.0}{\rmdefault}{\mddefault}{\updefault}{\color[rgb]{0,0,0}$c_4$}%
}}}}
\put(2489,614){\makebox(0,0)[lb]{\smash{{\SetFigFont{10}{12.0}{\rmdefault}{\mddefault}{\updefault}{\color[rgb]{0,0,0}$v'_1$}%
}}}}
\end{picture}%

\vspace{.5cm}

\noindent of maximum degree three and number of vertices bounded by $n^2$. It is known that there is a polynomial-time algorithm for embedding this graph into a grid of size quadratic in the number of vertices, thus obtaining the grid $\mathcal{G}'_\Phi:=$

\vspace{.5cm}
\hspace{1cm} \begin{picture}(0,0)%
\includegraphics{planargrid441.pstex}%
\end{picture}%
\setlength{\unitlength}{3947sp}%
\begingroup\makeatletter\ifx\SetFigFont\undefined%
\gdef\SetFigFont#1#2#3#4#5{%
  \reset@font\fontsize{#1}{#2pt}%
  \fontfamily{#3}\fontseries{#4}\fontshape{#5}%
  \selectfont}%
\fi\endgroup%
\begin{picture}(4276,2386)(225,-1771)
\put(964,220){\makebox(0,0)[lb]{\smash{{\SetFigFont{10}{12.0}{\rmdefault}{\mddefault}{\updefault}{\color[rgb]{0,0,0}$v_1$}%
}}}}
\put(1701,201){\makebox(0,0)[lb]{\smash{{\SetFigFont{10}{12.0}{\rmdefault}{\mddefault}{\updefault}{\color[rgb]{0,0,0}$v'_1$}%
}}}}
\put(951,-449){\makebox(0,0)[lb]{\smash{{\SetFigFont{10}{12.0}{\rmdefault}{\mddefault}{\updefault}{\color[rgb]{0,0,0}$c_2$}%
}}}}
\put(439,-461){\makebox(0,0)[lb]{\smash{{\SetFigFont{10}{12.0}{\rmdefault}{\mddefault}{\updefault}{\color[rgb]{0,0,0}$c_1$}%
}}}}
\put(1408,-430){\makebox(0,0)[lb]{\smash{{\SetFigFont{10}{12.0}{\rmdefault}{\mddefault}{\updefault}{\color[rgb]{0,0,0}$c_3$}%
}}}}
\put(3126,-411){\makebox(0,0)[lb]{\smash{{\SetFigFont{10}{12.0}{\rmdefault}{\mddefault}{\updefault}{\color[rgb]{0,0,0}$c_4$}%
}}}}
\put(4095,-405){\makebox(0,0)[lb]{\smash{{\SetFigFont{10}{12.0}{\rmdefault}{\mddefault}{\updefault}{\color[rgb]{0,0,0}$c_5$}%
}}}}
\put(1489,-1411){\makebox(0,0)[lb]{\smash{{\SetFigFont{10}{12.0}{\rmdefault}{\mddefault}{\updefault}{\color[rgb]{0,0,0}$v_2$}%
}}}}
\put(2939,-936){\makebox(0,0)[lb]{\smash{{\SetFigFont{10}{12.0}{\rmdefault}{\mddefault}{\updefault}{\color[rgb]{0,0,0}$v_3$}%
}}}}
\put(2401,-436){\makebox(0,0)[lb]{\smash{{\SetFigFont{10}{12.0}{\rmdefault}{\mddefault}{\updefault}{\color[rgb]{0,0,0}$v_4$}%
}}}}
\put(3139,214){\makebox(0,0)[lb]{\smash{{\SetFigFont{10}{12.0}{\rmdefault}{\mddefault}{\updefault}{\color[rgb]{0,0,0}$v''_1$}%
}}}}
\end{picture}%

\vspace{.5cm}

\noindent For technical reasons which will become clear, we double all the coordinates involved to generate the new grid, whose size remains quadratic in $n$, $\mathcal{G}_\Phi:=$ 

\vspace{.5cm}
\hspace{1cm} \includegraphics{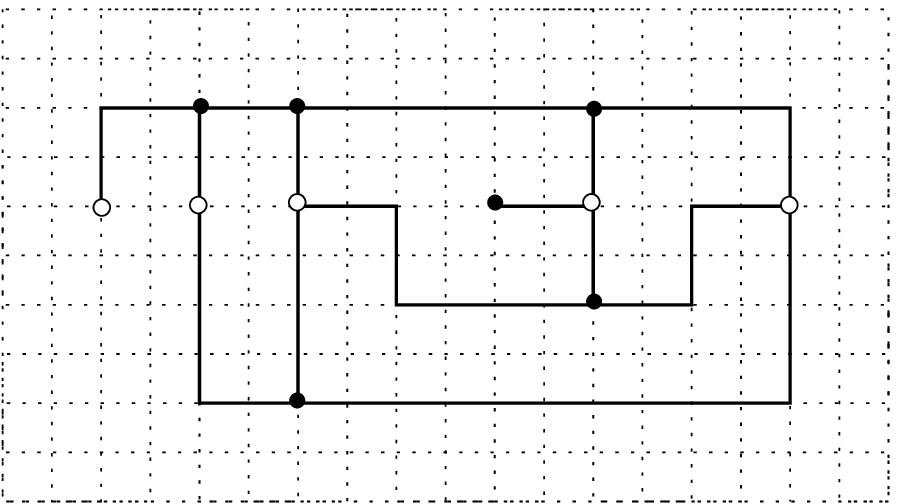}
\vspace{.5cm}

\noindent It remains now to translate this grid into a board $\mathscr{G}$. We will place $16 \times 16$ sub-boards over the intersections of each of the dotted lines of our grid. Where there is neither a vertex nor an edge passing through this intersection, we will place a Black Sub-board. The variable/subvariable vertices may appear, w.l.o.g, as Variable Sub-boards (if these have degree less than three, we need not worry since, from doubling the grid coordinates, any adjacent sub-board on that side will be black). The clause vertices may appear as Clause Sub-boards (again we need not worry if they have degree less than three). Where we have split a variable because it appeared more than three times in $\Phi$ we must join up the respective subvariable vertices to ensure that their truth values are consistent. We do this by connecting the $T$ points on the respective Variable Sub-boards. Similarly must we join variables with their positive occurrences within clauses. Again, we do this by connecting the $T$ point on the Variable Sub-board to the $T$ point on the relevant Clause Sub-board. Finally, we must join variables with their negative occurrences within clauses, ensuring that we reverse the polarity of true and false. We do this by connecting the $T$ point on the Variable Sub-board to the $F$ point on the relevant Clause Sub-board. In all cases we can do this with our assorted Join and Turn sub-boards with the necessary polarity changes and with the necessary male-female marriage. This is due to our doubling the coordinates of the grid, which ensured that between any turns or variables or clauses in the grid there must be a horizontal (respectively, vertical) distance of at least two. This ensures that on our board between Turn, Variable or Clause sub-boards there must be at least one male-male sub-board. Depending on which polarity is required, we may select either the Join MM or the Join-Switch MM.

We now have the board $\mathscr{G}:=$

\vspace{.5cm}
\hspace{1cm} \includegraphics{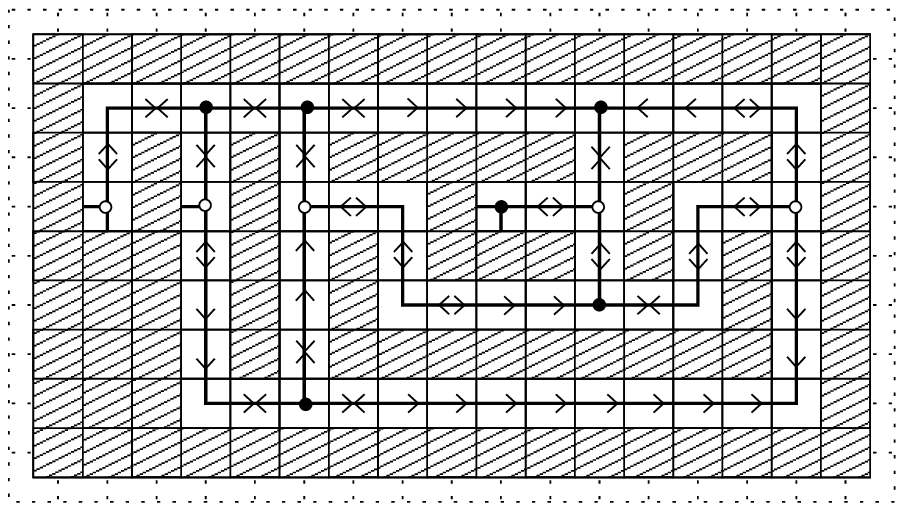}
\vspace{.5cm}

\noindent Note that the Turn, Variable and Clause sub-boards each possess certain intrinsic polarities relative to themselves. This is why, in our example, there is a Join-Switch MM between the Variable sub-boards representing the subvariables $v_1$ and $v'_1$, despite the fact that we want to ensure a consistent (identical) valuation on these. The reason for this is that the `left' side of these Variable Sub-boards intrinsically has the opposite polarity to the `right'. In our example, the board specified is $17.16 \times 9.16$. It has the additional properties that: all male-male sub-boards have been propagated so they only appear adjacent to Turn, Clause or Variable sub-boards; and all Join-Switch MM Sub-boards have been propagated so that they only appear adjacent to Variable Sub-boards. In the enlargement of the board $\mathscr{G}$ given in Figure~1, the intended truth labelling of the devices is given. This enables us to read the paths inbetween subvariables, and between subvariables and clauses, to verify that the polarities of the original instance have been respected.

\begin{figure}
\hspace{1cm} \includegraphics{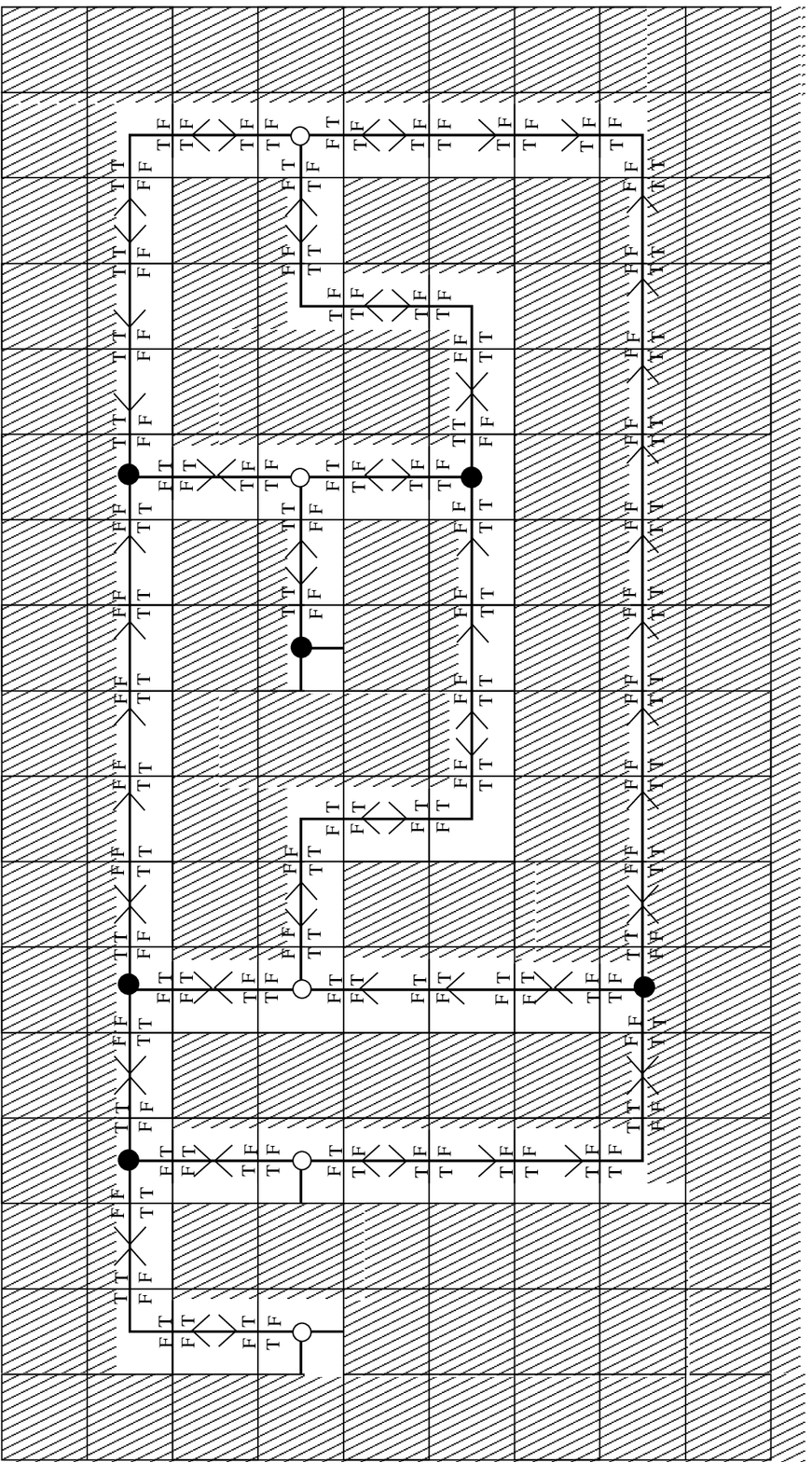}
\caption{Enlargement of the board $\mathscr{G}$ obtained in our reduction. The intended truth labelling is given, although it should be borne in mind that this plays no part in the board physical.}
\end{figure}

\noindent Finally, we must establish how many of each of the sundry sub-boards we have used, that we may sum up the tariff of queens according to the tariffs of each sub-board as given in the lemmas of the previous section. This tariff $m$ will always be bounded by the total size of the board $\mathscr{G}$. In the case of our example we may calculate thus:
\[
\begin{array}{llllr}
\mbox{Clause sub-boards} & 5 & \mbox{@} & 21 & 105 \\
\mbox{Variable sub-boards} & 6 & \mbox{@} & 20 & 120 \\
\mbox{Turn sub-boards} & 8 & \mbox{@} & 12 & 96 \\
\mbox{Join FM sub-boards} & 22 & \mbox{@} & 16 & 352 \\
\mbox{Join MM sub-boards} & 12 & \mbox{@} & 17 & 204 \\
\mbox{Join-Switch MM sub-boards} & 10 & \mbox{@} & 16 & 160 \\
& & m & = & 1037. \\
\end{array}
\]

\end{proof}

\section{The boundary of tractability}

Having obtained evidence of intractability, here in the form of \NP-completeness, it is natural to ask what manner of restriction may be imposed on the problem in order to guarantee polynomial-time tractability. In general, we have allowed a queen unlimited movement along rows, columns and diagonals. Might the problem become simpler to solve if we were to restrict this range? A perusal of our gadgets reveals that the problem remains \NP-complete, under the same proof, even if a queen is restricted to move only two squares in any direction. This then raises the question of the complexity of the problem if the queens were restricted to move only one square in any direction -- i.e. were restricted to move like kings. It is straightforward to see that all of our sub-board gadgets work for kings, except the clause gadget. In fact we can generate a clause gadget that does work for kings, shown in Figure~2, and we may thus derive that this problem, \textsc{Kings}, is also \NP-complete.

\begin{figure}
\hspace{3.5cm}  \includegraphics{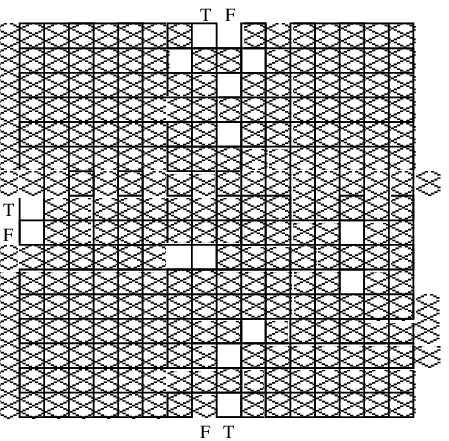}
\caption{This $16 \times 16$ sub-board has the property that it can legitimately hold at most $25$ kings, and this is achievable iff one king is placed in each of the three exterior double squares and not all of the three squares labelled $F$ contain kings.}
\end{figure}

That the problem \QP\ becomes polynomial-time tractable for a prior fixed tariff $m$ is straightforward; whether the problem is fixed-parameter tractable in the tariff $m$ is an interesting open question.

\section{Acknowledgements}

The author would like to thank Florent Madelaine for posing the question as to what cousin of the classical $n$-queens problem might be \NP-complete. The author would also like to thank Stefan Dantchev for his forbearance on the topic.

\bibliographystyle{splncs}
\bibliography{QueensProblem2}

\end{document}